\documentclass[12,preprint]{aastex}
\begin{document}


\title{Hall instability of thin weakly-ionized stratified  Keplerian disks}


\author{ Yuri M. Shtemler, Michael Mond and Edward Liverts }
\affil{Department of Mechanical Engineering,  Ben-Gurion
University of the Negev, \\ P.O. Box 653, Beer-Sheva 84105,
Israel}

\email{shtemler@bgu.ac.il}

\begin{abstract}
The stratification-driven Hall instability in a weakly ionized
polytropic plasma  is investigated in the local approximation
within an equilibrium Keplerian disk of a small aspect ratio
$\epsilon$. The leading order of the asymptotic expansions in
$\epsilon$  is  applied to  both equilibrium and perturbation
problems.  The equilibrium disk with an embedded purely toroidal
magnetic field is found to be stable to radial, and unstable to
vertical short-wave perturbations. The marginal stability surface
was found in the space of the local Hall and inverse plasma beta
parameters, as well as the free parameter of the
 model related to the total current. To estimate the  minimal values of the
equilibrium  magnetic field that leads to the instability, it was
constructed as a sum of a current free  magnetic field  and the
simplest approximation for magnetic field created by the
distributed electric current.
\end{abstract}


\section{Introduction}\label{sec:intro}

The dynamics of thin rotating gaseous disks under the influence of
a magnetic field is of great importance to numerous astrophysical
phenomena. Of particular importance is the study of the various
instabilities that may arise within the  disks whose inverse
growth rates are comparable to the rotation period. The
rediscovery of the magneto-rotational instability (MRI) within
astrophysical context (Balbus and Hawley 1991) gave the sign to a
series of extensive investigations of hydromagnetic instabilities
of rotating disks.

While most of the research has focused on the magnetohydrodynamics
(MHD) description of MRIs, the importance of the Hall electric
field to such astrophysical objects as protoplanetary disks has
been pointed out by Wardle (1999). Since then, such works as
Balbus and Terquem (2001), Salmeron and Wardle (2003, 2005), Desch
(2004), Urpin and Rudiger (2005), and Rudiger and Kitchatinov
(2005) have shed light on the role of the Hall effect in the
modification of the MRIs. However, in addition to modifying the
MRIs, the Hall electromotive force gives rise to a new family of
instabilities
 in a density-stratified environment. That new
family of instabilities is characterized by the Hall parameter
$\hat{\xi}$ of the order of unit ($\hat{\xi}$ is the ratio of the
Hall drift velocity, $V_{HD}$, to the characteristic velocity,
where $V_{HD}=l_i^2 \Omega_i / H_*$, $l_i$ and $\Omega_i$ are the
inertial length and Larmor frequency of the ions, and  $H_*$ is
the density inhomogeneity length, see  Huba (1991), Liverts and
Mond (2004), Shtemler and Mond (2006)). As  shown by Liverts and
Mond (2004) and by Kolberg et al. (2005), the Hall electric field
combined with density stratification, gives rise to two modes of
wave propagation. The first one is the stable fast magnetic
penetration mode, while the second is a slow quasi-electrostatic
mode that may become unstable for short enough density
inhomogeneity length  $H_*$. The latter is termed the Hall
instability, and the aim of the present work is to introduce it
within astrophysical context.

Although the magnetic field configuration in real disks is largely
unknown, observations and numerical simulations indicate that the
toroidal magnetic component  may be of the same order of magnitude
as, or some times dominate the poloidal  field due to the
differential rotation of the disk, even if the magnetic field is
created continuously from an initial purely poloidal field (see
e.g. Terquem and Papaloizou 1996, Papaloizou and Terquem 1997,
Hawley and Krolik 2002, Proga 2003).

The present study of the Hall instability is carried out for thin
equilibrium Keplerian disks with embedded purely toroidal magnetic
fields. While the accepted approach to the steady state of the
rotating disks in most astrophysics-related hydromagnetic
stability researches is that of a "cylindrical disk", a more
realistic model of thin disks is adopted in the current work.
Thus, an asymptotic approach  developed for equilibrium
non-magnetized rotating disks (Regev, 1983, Klu\'zniak and Kita
2000, Umurhan et al 2006), is employed here in order to construct
a family of steady-state solutions for the rotating disks in a
toroidal magnetic field within the Hall-MHD model, and to their
further study of their stability. It is shown that in such
radially stratified systems the Hall instability indeed occurs
with the inverse growth rate of the order of the rotation period.

The paper is organized as follows. The dimensionless governing
equations are presented in the next Section. In Section 3 the
Hall-MHD equilibrium configuration of a  rotating disk subjected
to a gravitational potential of a central body and a toroidal
magnetic field, is described in the thin-disk limit. Linear
analysis of the Hall stability of toroidal
 magnetic fields in thin equilibrium
Keplerian disks and the results of  calculations for both
arbitrary and specific magnetic configurations are presented
 in Section 4. Summary and conclusions are given
in Section 5.

\section{The physical model. Basic equations}
\subsection{Dimensional basic equations}
The stability of radially and axially stratified thin rotating
disks threaded by a toroidal magnetic field under short-wave
perturbations is considered. Viscosity and radiation effects are
ignored. The Hall electric field is taken into account in the
generalized Ohm's law which is derived from the momentum equation
for the electrons fluid  neglecting the electron inertia and
pressure. Under the assumptions mentioned above, the dynamical
equations that describe the partially ionized plasmas are:
\begin{mathletters}
\begin{equation}
\frac{\partial n} {\partial t}
               + \nabla \cdot(n {\bf{V}} ) =0,\\
\end{equation}
\begin{equation}
m_{i}n \frac{ D \bf {V} }{Dt} = -\nabla P+\frac{1}{c} {\bf{j}}
\times {\bf{B}}- m_{i} n \nabla \Phi
,\\
\end{equation}
\begin{equation}
\frac{D  } {D t}(P n^{-\gamma})=0,\\
\end{equation}
\begin{equation}
\frac{\partial {\bf{B} }} {\partial t}+ c \nabla \times
 {\bf{E}}=0,\,\nabla \cdot {\bf{B}}=0,
\end{equation}
\begin{equation}
 {\bf {E} }=  -  \frac{1}{c}{\bf{V}} \times {\bf{B}} +
\frac{1}{ec} \frac{\bf{j}
  \times \bf{B} }{n_{e} },\,\,\,{\bf {j}} =
 \frac{c}{4 {\pi}}
 \nabla \times {\bf{B}}.\,\,\,\,
\end{equation}
\end{mathletters}
Cylindrical coordinates $\{r, \theta,z\}$  are adopted throughout
the paper with the associated unit basic vectors $\{ \mathbf{i}_r,
\mathbf{i}_\theta, \mathbf{i}_z \}$;   $t$ is time;  $\Phi$ is the
gravitational potential of the central object; $\Phi(R)=-GM/R$;
$R^2=r^2+z^2$; $G$ is the gravitational constant; and $M$
 is the total mass of the central object;  $c$  is the speed of light;  $e$ is the electron charge;
$\mathbf{E}$ and $\mathbf{B}$
  are the  Hall electric and magnetic fields; $\mathbf{j}$
is the electrical current density;
 $\mathbf{V}$  is the plasma velocity;
  $D/Dt=\partial/\partial t+(\mathbf{V} \cdot\nabla)$ is the material
  derivative; $n$ is the number density;
 $P=P_e+P_i+P_n$  is the total plasma pressure;
   $P_k$   and   $m_k$  are the species pressures and masses ($k=e,i,n $);
   subscripts $e$, $i$ and $n$ denote the electrons, ions and neutrals.
In compliance with the observation data for protoplanetary disks,
the density decreases outward the center of the disk according to
the law $n \sim 1/r^\sigma$ with the density exponent $\sigma\approx 2.75$ \cite{GiuDut}.
In a standard polytropic model of plasmas $\sigma = 3/(\gamma
-1)$, this yields  the specific heat ratio
 $\gamma \approx 2.1$ that is somewhat higher than the standard value $\gamma =5/3$ used
here.
    Since the plasma is assumed to be quasi-neutral and weakly ionized
\begin{equation}
n_e \approx n_i \approx \alpha n_n , n \approx n_n.
\end{equation}
This indicates the increasing role of the Hall effect in the
generalized Ohm law with vanishing ionization degree,
$\alpha=n_e/(n_e+n_n)<<1$.

\subsection{Scaling procedure}

The physical variables are now transformed into a dimensionless
form:
\begin{equation}
f_{nd}=f / f_*,
\end{equation}
where  $f$   and  $f_{nd}$ stand for any physical dimensional and
non-dimensional variables, while the characteristic scales $f_*$
are defined as follows:
\begin{mathletters}
\begin{equation}
V_*=\Omega_* r_* ,\,\,t_*=\frac{1}{\Omega_*},\,\,
\Phi_*={V_*}^2,\,\, m_*= m_i,\,\,
n_*=n_n,\,\,\,
\end{equation}
\begin{equation}
 P_*=K(m_* n_*)^\gamma,\,\,
\,\,j_*=\frac{c}{4\pi}\frac{B_*}{r_*},\,\, E_*=\frac{V_*
B_*}{c}.
\end{equation}
\end{mathletters}
Here $\Omega_*=(GM/r^3)^{1/2}$  is the value of the Keplerian
angular velocity of fluid at the characteristic radius $r_*$;
 $K$ is the dimensional constant in the steady-state dimensional
polytropic law. The characteristic values of the magnetic field
$B_*$ and radius $r_*$ are the dimensional parameters of the
problem. This yields the following dimensionless system (omitting
the subscript $nd$ in non-dimensional variables):
\begin{mathletters}
\begin{equation}
\frac{\partial n} {\partial t}
               + \nabla \cdot(n {\bf{V}} ) =0,\,\,\,\,\,\,\,\,\,\,\,\,\,\,
\end{equation}
\begin{equation}
n \frac{ D \bf {V} }{Dt} = -\frac{1}{M_S^2}\nabla P +
\frac{1}{\beta M_S^2} {\bf{j}} \times {\bf{B}}-  n \nabla
\Phi,\,\,\,\Phi(r,z)=-{1 \over (r^2+z^2)^{1/2}},
\end{equation}
\begin{equation}
\frac{D  } {D t}(P n^{-\gamma})=0,
\end{equation}
\begin{equation}
\frac{\partial {\bf{B} }} {\partial t}+  \nabla \times
 {\bf{E}}=0,\,\nabla \cdot {\bf{B}}=0,
\end{equation}
\begin{equation}
 {\bf {E} }=  -  {\bf{V}} \times {\bf{B}} + \xi \frac {\bf{j}
\times \bf{B} }{n},\,\,\,\,{\bf {j}} =
  \nabla \times {\bf{B}},
\end{equation}
\end{mathletters}
where  $M_S$ and $\beta$ are the  Mach number and plasma beta, and
$\xi$ is the Hall coefficient:
\begin{equation}
M_S=\frac{V_*}{c_{S*}},\,\,\,\beta=4\pi \frac{P_*}{B_*^2},\,\,\,
\xi=\frac{\Omega_i}{ \Omega_*}\bigg{(}
\frac{l_i}{r_*}\bigg{)}^2\equiv \frac{B_* c}{4 \pi e \alpha n_*
\Omega_*r^2_*},
\end{equation}
$c_{S*}=\sqrt{\gamma P_*/n_*}$, $l_i=c/\omega_{pi}$  and
$\Omega_{i}=eB_*/(m_i c)$ are the inertial length and the Larmor
frequency of  ions, respectively, $\omega_{pi}=\sqrt{4\pi e^2
n_i/m_i}$ is the plasma frequency of the ions.

A common property of thin Keplerian disks is their highly
compressible motion with large Mach numbers $M_S$
\begin{equation}
\frac{1}{M_S}=\epsilon<<1,\,\,\,\epsilon=\frac{H_*}{r_*},
\end{equation}
where $\epsilon$ is the aspect ratio of the disk; $H_*$ is a
characteristic dimensional  thickness of the disk. It is
convenient therefore to use a stretched vertical coordinate
\begin{equation}
\zeta=\frac{z}{\epsilon},\,\,\,h =\frac{H}{\epsilon}.
\end{equation}

\section{Equilibrium solution in thin-disk limit for toroidal magnetic
configuration}

The steady-state equilibrium of rotating magnetized disks  may now
be obtained to leading order in $\epsilon$ by writing all physical
quantities as asymptotic expansions in small $\epsilon$ (similar
to Regev, 1983, Klu\'zniak and Kita 2000, Umurhan et al 2006 in
their analysis of accretion disks, see also references therein).
This yields for the gravitational potential in the Keplerian
portion of the disk:
\begin{equation}
\Phi(r,\zeta)=-\frac{1}{r}+\epsilon^2\frac{\zeta^2}{2r^3}+O(\epsilon^4).
\end{equation}
To zeroth order in   $\epsilon$ , the toroidal velocity
$V_{\theta}$ is described by the Keplerian law:
\begin{equation}
V_{\theta}=\Omega(r)r,\,\,\, \Omega(r)=r^{-3/2},
\end{equation}
the rest of  the velocity components  are of  higher order in
$\epsilon$,
  and their input into the stability analysis may thus be neglected.
Let us assume now that the magnetic field is purely toroidal and
depends on both $r$  and   $z$:
\begin{mathletters}
\begin{equation}
{\bf{B}}=B_\theta(r,z) {\bf{i}}_{\theta},\,\,\,{\bf{j}}=
{\hat\nabla}\times {\bf{B}},
\end{equation}
\begin{equation}
{\hat \nabla}={\bf{i}}_r{\partial\over \partial r}
+\frac{1}{\epsilon}{\bf{i}}_z\frac{\partial}{\partial
\zeta},\,\,\,{\bf{j}}=\{j_r,0,j_z\}= \{-\frac{1}{\epsilon}
{\partial {B_\theta}\over \partial \zeta}, 0,\frac{1}{r}
{\partial(r B_\theta)\over \partial r}\}.
\end{equation}
\end{mathletters}

Since the vector product of the two toroidal vectors $\bf{V}$ and
$\bf{B}$ is zero, it follows from  system (1)  that the Hall
electric field has  the potential $\phi$:
\begin{mathletters}
\begin{equation}
\frac{\bf{j}\times \bf{B}}{n}={\hat \nabla}\phi
\end{equation}
\mbox{or, equivalently,}
\begin{equation}
n {\partial \phi \over \partial r}=-{B_\theta \over r}{\partial
(rB_\theta) \over \partial r},\,\,\, n {\partial \phi \over
\partial \zeta}=-B_\theta {\partial B_\theta \over \partial\zeta}.
\end{equation}
\end{mathletters}

Substituting Eqs. (12) into the vertical component of the momentum
equation (5b) yields  in  the leading order in $\epsilon$:
\begin{equation}
c_S=\sqrt{\gamma\frac{P}{n}},\,\,\,\, P=n^\gamma,\,\,\, n=\nu
\bigg [ {\phi (r,\zeta)\over \beta}+{h^2-\zeta^2 \over r^3}
\bigg]^{1 \over \gamma -1},\,\,\,\, \nu=\bigg{(}{\gamma-1 \over
2\gamma}\bigg{)}^{1 \over \gamma -1}.
\end{equation}
Here $h$ is the disk thickness, where $n$ and  $\phi$ equal zero
\begin{equation}
n=0,\,\,\,\,\phi=0\,\,\, \mbox{at}\,\,\, \zeta=h,
\end{equation}
 $h(r)$ is scaled so that $h(1)=1$;   $h(r)$ should be specified
 to close the problem (see Section 4).

 Finally, a general expression for the toroidal magnetic field as well as for the
 Hall potential may be derived by solving the induction equation that
 assumes the following form:
\begin{equation}
{\partial (r^2n) \over \partial \zeta} {\partial (r^2 B_\theta^2)
\over \partial r} - {\partial (r^2n) \over \partial r} {\partial
(r^2 B_\theta^2) \over \partial \zeta}=0.
\end{equation}
This equation  is satisfied for  $r B_\theta$ that is an arbitrary
function of  $N$ (Kadomtzev 1976), where  $N$  has the meaning of
the inertial moment density (see additionally  Section 4.3):
\begin{equation}
 B_\theta= r^{-1}I(N),\,\,\,N=r^2n.
\end{equation}
 This yields for the Hall  potential
\begin{equation}
\phi(N) =-\int_{0}^{N}{I(N) \dot{I}(N)\over N} d N,
\end{equation}
where the arbitrary constant in the potential is chosen from the
condition that the potential is zero at the disk edge where $n=0$,
i.e. $\phi(0)=0$; the upper dot denotes a derivative with respect
to the argument. If the arbitrary function $I(N)$ is given, then
Eqs. (13) and (17) constitute implicit algebraic equations for the
number density $n$ and the  potential $\phi$.

\section{Hall instability of thin Keplerian  disks}
The stability properties of the axially symmetric steady-state
equilibrium solutions for thin Keplerian disks are investigated
now under small axially symmetric perturbations, and find the
dispersion relation in the short wave limit.

\subsection{Thin disk
approximation for perturbed problem }

We start by linearizing the Hall MHD equations (5) about the
equilibrium solution within the Keplerian portion of the thin
disk. For that purpose, the perturbed variables are given by:
\begin{equation}
{\overline F}(r,z,t)= F(r,\zeta)+ F'(r,\zeta,t).
\end{equation}
Here  $\overline F$  stands for any of the physical variables, $F$
denotes the equilibrium value, while   $F'$ denotes the
perturbations  of the following form:
\begin{equation}
{\bf{B'}}(r,z)=B'_\theta {\bf{i_\theta}},\,\,\,{\bf{j'}} \approx
\{{1 \over \epsilon}{\partial B'_\theta \over
\partial \zeta}, 0,{\partial B'_\theta \over
\partial r}\},\,\,\,
{\bf{V}}'=\{0,0,V'_z\}.
\end{equation}

Inserting  the relations (18)-(19) into Eqs. (5), linearizing them
 about the equilibrium solution and
applying  the principle of the least possible degeneracy of the
problem \cite{VanDyke} to the result, we obtain that $B'_\theta $
is of the order  $\epsilon^0$, while the axial perturbed velocity
$V'_z$ is of the order $\epsilon^1$ and  should be rescaled.
Furthermore, we rescale  the Hall parameter, keeping in mind that
it should be of the order of $\epsilon^1$, since its higher and
lower limits in $\epsilon$ lead to  more degenerate problems.
Thus, rescaling the axial perturbed velocity and Hall parameter,
we derive
\begin{equation}
{\hat V}'_z=\frac{V'_z}{\epsilon},\,\,\,{\hat
\xi}=\frac{\xi}{\epsilon} \equiv \frac{\Omega_i l_i^2}{ \Omega_*
r_* H_*}.\,\,\,
\end{equation}
Here the rescaled Hall parameter ${\hat \xi}$ is equal to the
ratio of the Hall drift velocity, $V_{HD}$, to the characteristic
velocity, $V_{*}=\Omega_* r_*$, and the characteristic disk
thickness $H_*= \epsilon r_*$ (see Eq. (7)) is the density
inhomogeneity length. It is noted that the limit ${\hat \xi}<<1$
yields the MHD model. Finally this leads to the following
linearized system in the leading order in $\epsilon$:
\begin{mathletters}
\begin{equation}
{\partial n' \over \partial  t} + n {\partial {\hat V}'_z \over
\partial \zeta} + {\partial n \over
\partial  \zeta}{\hat V}'_z=0,
\end{equation}
\begin{equation}
n{\partial {\hat V}'_z  \over \partial  t}=-{\partial P' \over
\partial \zeta} -{1 \over \beta} {\partial B_\theta \over \partial
\zeta}B'_\theta -{1 \over \beta} B_\theta{\partial B'_\theta \over
\partial \zeta},
\end{equation}
\begin{equation}
{\partial P' \over \partial  t}-c_S^2{\partial n' \over \partial
t}=0,\,\,\,
\end{equation}
\begin{equation}
{\partial B'_\theta \over \partial  t}+ {\partial (B'_\theta {\hat
V}'_z ) \over \partial \zeta}+ {\hat \xi}{B_\theta \over n^2}
\bigg \{ {1 \over r}{\partial (r B_\theta) \over \partial
r}{\partial n'
 \over \partial  \zeta}-{1
\over r^2}{\partial (r^2 n) \over \partial  r}{\partial B'_\theta
 \over \partial  \zeta}+{\partial  n \over \partial \zeta}{\partial B'_\theta
 \over \partial  r}
 -{\partial  B_\theta \over \partial \zeta}{\partial n'
 \over \partial  r}\bigg \}
+{\hat \xi} \Psi=0.
\end{equation}
\end{mathletters}
Here the perturbed toroidal magnetic field identically satisfies
the relation $\nabla \cdot {\bf{B}}'=0$; $\Psi$ denotes the terms
in the magnetic induction equation that do not include derivatives
of perturbations. The latter will be neglected below in the short
wave limit, and it is not presented here for brevity.

\subsection{Dispersion relation in the short wave limit}

 The perturbed quantities are assumed to satisfy the following
conditions:
\begin{equation}
\mid \frac{1}{ F'}{\partial F' \over \partial t} \mid \sim \mid{1
\over F'}{\partial F' \over \partial\zeta}\mid\sim \mid {1 \over
F'}{\partial F' \over \partial r}\mid \,\,\gtrsim \,\,\mid {1
\over F}{\partial F \over \partial \zeta}\mid\sim\mid{1 \over
F}{\partial F \over
\partial r} \mid \sim \frac{1}{r}\sim 1.
\end{equation}
This allows us to assume the following local approximation for the
perturbations:
\begin{equation}
F'(r,\zeta,t)=f' \exp(-i\omega t+i { k}_r r+i {k}_\zeta \zeta).
\end{equation}
Here $f'$
 denotes the amplitude of fluctuations about the
equilibrium; $\omega$ is the (generally complex) frequency of
perturbations;  ${\bf {\hat k}}=\{{ k}_r,  { k}_\zeta \}$ is the
wave vector in the scaled coordinates $\{r,\zeta\}$. It should be
distinguished from the wave vector ${\bf k}=\{k_r, k_z\}$ in the
$\{r,z\}$ coordinates,   $k_z={ k}_\zeta /\epsilon$. Condition
(22) may now be written as
\begin{equation}
 \omega \sim { k}_\zeta\sim { k}_r\,\,\gtrsim\,\, \mid {1
\over F}{\partial F \over
\partial \zeta}\mid\sim\mid{1 \over F}{\partial F \over
\partial r} \mid \sim \frac{1}{r}\sim 1.
\end{equation}

Using relations (23)-(24), the system (21) may be presented as the
following system of homogeneous linear algebraic equations:
\begin{mathletters}
\begin{equation}
-{\hat C} n'+n  {\hat V}'_z=0,
\end{equation}
\begin{equation}
-{\hat C} n  {\hat V}'_z=- P' - \beta^{-1} B_\theta
B'_\theta,\,\,\,\,V'_r=0,
\end{equation}
\begin{equation}
P'=c_S^2n',\,\,\,\,
\end{equation}
\begin{equation}
{\hat C} B'_\theta+B_\theta  {\hat V}'_z+{\hat \xi}{B_\theta \over
n^2}\bigg {\{} {1\over r}{\partial (r B_\theta) \over
\partial \rho} n'
 -{1\over r^2}{\partial (r^2 n) \over \partial \rho}
 B'\bigg{\}}=0.\,\,\,
\end{equation}
\end{mathletters}
Here  the scaled phase velocity ${\hat C}=\omega/{k}_\zeta$ in the
scaled coordinates $\{r,\zeta\}$  is related to the phase velocity
$C$ in the cylindrical coordinates $\{r,z\}$ as  ${\hat C}=C /
\epsilon$, and
\begin{equation}
l_\zeta{\partial  \over \partial \rho}=l_\zeta{\partial
 \over \partial r}  -
 l_r {\partial  \over \partial \zeta},\,\,\,
l_r=\frac{{k}_r}{{\hat k}},\,\,\,\,l_\zeta=\frac{{k}_\zeta}{{\hat
k}},\,\,\,\,{\hat k}=\sqrt{{ k}_r^2+{k}_\zeta^2}.
\end{equation}
Note that since   the radial wave number ${ k}_r$  is contained
only in the Hall term in the magnetic induction Eq. (25d), in the
leading order in $\epsilon$ the standard MHD problem is free from
${\hat k}_r$.

The system of equations (25) has non-trivial solutions if the
following cubic eigenvalue equation for the local value of the
rescaled phase velocity ${ c}$ is satisfied:
\begin{equation}
X{ c}^3+ X { c}^2-  { c} - q =0.
\end{equation}
Parameters $X$ and $q$ are expressed through the local values of
the plasma beta $b$, Hall coefficient $x$ and new
 parameter $g$ whose the physical meaning  will be elucidated later
\begin{mathletters}
\begin{equation}
{ c}={{\hat C} \over x c_S},\,\,\,\ X={x^2 \over 1+b^{-1}} ,\,\,\,
q={1+g\,b^{-1} \over 1+b^{-1}},
\end{equation}
\begin{equation}
 b^{-1}= \beta^{-1}{B_\theta^2 \over c_S^2
n},\,\,\,\, x={\hat \xi}{B_\theta \over c_S n^2}{1\over
r^2}{\partial N \over
\partial \rho},\,\,\,\
g={ d\, ln\, I(N)\over d\,\, ln N},
\end{equation}
\end{mathletters}
where $B_\theta$ and $N$ are determined in Eqs. (16).
 Straightforward analysis of dispersion relation (27) (such
dispersion relation for the Hall instability of plasma in  slab
geometry has been derived and investigated in \cite{BrMor})
reveals that the system is stable for $0 \leq q\leq 1$, while the
instability may occur for the following two regimes:
\begin{equation}
1) \,\,\,q>1\,\,\, \mbox{and}
\,\,\,X^{-}<X<X^{+}\,\,\,\,\,\,\,\mbox{or} \,\,\,\,\,\,\, 2)
\,\,\,q<0 \,\,\,\mbox{and}\,\,\, X>X^{-}.
\end{equation}
Here the first instability regime  corresponds to  $g>1$, and the
second one to  $g<-b$ (see  Eqs. (28a)); $X^{-}$ and $X^{+}$ are
the roots of the following quadratic equation:
\begin{equation}
4qX^2+(1+18q-27q^2)X+4=0.
\end{equation}
 Although Eq. (30) provides for the dependence of
the  local Hall parameter $X$ vs $q$, it is more convenient to use
$X$ as a function of the local plasma beta $b$ and the free
parameter $g$, which may be related to the  natural parameter of
the disk - total current through the disk (see Subsection 4.3). As
a first result, it is noticed that the current free magnetic
configuration with $g\equiv 0$ is stable since $0<q<1$. In
addition, it is noted that the quantities $X$ and $q$ (or,
equivalently, $X$, $g$ and $b$) which determine the stability
properties of the disk, depend on the
 local values of
the equilibrium number density and magnetic field which are
determined in Section 3 up to  the free function $I(N)$  of the
inertial moment density.

Figure 1 depicts the family of the marginal stability curves which
correspond to the real roots of  Eq. (30). These curves are
presented in the plane $X - b^{-1}$ of the local Hall parameter
and inverse plasma beta with a single parameter of the family,
$g$. According to the relations (29), there are two instability
regimes (Figs. 1a and 1b). Thus, in Fig. 1a the values of $X$
located between the  upper and bottom branches of the marginal
stability curves give rise to the first regime of instability,
while the disk is stable if $X$ falls outside that interval. In
the limit $b^{-1} \to 0$, we have $q \to 1$, and Eq. (30) has a
double root $X=1$. This means that in the limit of high plasma
beta the instability interval becomes very narrow  and the disk is
stable. In Fig. 1b, the values of $X$ located higher the line
$X^{-}$
 correspond to the second instability regime in (29) ($q<0$ mode),
 and the disk is stable otherwise.

\subsection{Model equilibrium solutions for toroidal magnetic fields}

The marginal stability curves in Fig. 1 provide for only a
qualitative description of the disk stability, since they are
depicted in terms of the local parameters  which  are determined
up to the free function $I(N)$. To estimate qualitatively the
stability characteristics of the system  a specific example of a
toroidal magnetic field is considered. The free function $I(N)$ is
presented as a sum of the current free magnetic field (created by
a current localized along the disk axis), and the simplest
quadratic approximation for the magnetic field created by an
electric current distributed over Keplerian portion of the disk:
\begin{equation}
I(N)=I_0+I_2{N^2 \over n_0^2},\,\,\,N =r^2 n,
\end{equation}
where  $I_0$,  $I_2$   are constant coefficients; $n_0=n(1,0)$ is
the value of number density at the reference point in the midplane
$\{r=1,\, z=0\}$ (analogous notations will be used below for the
Hall electric potential $\phi_0=\phi(1,0)$, free parameter
$g_{0}=g(1,0)$,   Hall coefficient ${\hat \xi}_0={\hat \xi}(1,0)$,
 etc). The absence of a linear term in $N$ in Eq. (31) is a
result of the requirement that the Hall electric potential $\phi$
is finite at the disk edge $z=h$, which in turn allows  the number
density  to be zero.

Substituting Eq. (31) into Eq. (17)
and using  the normalization condition $B_{\theta}=1$ at the
reference point 
$\{r=1,\, z=0\}$ yield the following expression for the Hall
electric potential:
\begin{equation}
\phi(N)=-2 I_2{N \over n_0^2}\bigg ( I_0+{I_2 \over 3}{N^2 \over
n_0^2}\bigg ),\,\,\,I_0 +I_2=1.
\end{equation}
Definitions of $g$ and $I$ in Eqs. (28) and (31) relate the values
of  $I_2$,  $g_0$ and the total current through the disk $I_t$:
\begin{equation}
g_0=2 I_2, \,\,\, I_t=2 \pi \bigg[ 1- \bigg(1- {N_\infty^2 \over
n_0^2}\bigg){ g_0 \over 2}\bigg],
\end{equation}
where  $N_\infty=(r^2 n_m)_{r\to \infty}$  is  constant (see Figs.
2 and 3). This leaves $ g_0$ or, equivalently, the natural
parameter of the disk - total current $I_t$ as the single free
parameter of the model. Equations (13) and (32)-(33)  provide for
the magnetic configuration in the equilibrium Keplerian disk.  To
overcome the presence of the arbitrary semi-thickness $h(r)$ in
relations (13), the small slope of the disk edge is assumed (that
is supported by the observation data for some protoplanetary disks
\cite{Calvet}), such that
 $h(r)\equiv 1$.  This yields  at the midplane:
\begin{equation}
n_m(r)=\nu\bigg(\beta^{-1}\phi_m+{1 \over r^3}\bigg)^{{1 \over
 \gamma -1}},\,\,\,
 \phi_m(r)=- {g_0 r^2 n_m \over n_0^2} \bigg[\bigg( 1-{g_0 \over
2}\bigg)+{g_0 \over 6}\bigg( {r^2 n_m \over n_0}\bigg)^2\bigg].
\end{equation}
Equations (34) form a coupled set of nonlinear equations for $n_m$
and $\phi_m$, while $n_0$  is determined from the following two
nonlinear equations for $n_0 \equiv n_m(1)$ and $\phi_0\equiv
\phi_m(1)$:
\begin{equation}
 n_0=\nu \bigg(\beta^{-1}\phi_0+1\bigg)^{{1 \over
 \gamma -1}},\,\,\,\phi_0=- {g_0 \over n_0} \bigg( 1-{g_0 \over
 3}\bigg).
\end{equation}
Numerical solution of Eqs. (34) and (35) reveals two main types of
equilibria that correspond to two possible signs of $g_0$, i.e.
signs of the contributions of the distributed current to the total
current. The midplane equilibrium magnetic field $B_{\theta m}$,
number density $n_m$, distributed electric current $j_{zm}$ and
inertial moment density $N_m$, are presented in Figs. 2 and 3 for
positive and negative $g_0$, respectively.
 In both cases the
equilibria are characterized by  a distributed electric current
localized at some radius that is moving outwards the disk center
with increasing plasma beta. In addition, while for positive $g_0$
both the equilibrium number density and magnetic field are
decreasing monotonically to zero at a finite plasma beta,
equilibria with negative $g_0$ exhibit a maximum in the number
density that corresponds to rings of denser material, as well as a
maximum in the magnetic field profile. Finally,  for $g_0<0$ the
equilibrium solutions exist only for inverse plasma beta larger
than a critical value corresponding to the vertical slope
($\beta^{-1}> 0.06$ in Fig. 3), otherwise the equilibrium solution
cannot be described by a single-valued function. Furthermore,
Figs. 2d and 3d demonstrate that the inertial moment density $N_m$
tends to a non-zero constant at infinity through a local minimum
(critical point) where $\partial_r N_m$=0.

\subsection{Hall instability of the model equilibrium}

Using the values of the free function $I(N)$ in the magnetic field
determined in the previous subsection, the family of the marginal
stability curves presented in Fig. 1 may be recalculated in terms
of the Hall coefficient ${\hat\xi}$, the inverse plasma beta
$\beta^{-1}$ and the free parameter of the model $g_0$ (i.e. the
parameters which are expressed through the input characteristics
values of the equilibrium disk, see Figs. 4, 5).

The values of ${\hat\xi}$ corresponding to the instability (see
relations (29)) lie either between the branches ${\hat\xi}^{-}$
and ${\hat\xi}^{+}$ for $q>1$
 (Fig. 4a, b for $g>0$) or above the branch ${\hat\xi}^{-}$  for
$q<0$   (Fig. 5a, b for $g<0$). Then  according to the data in
Figs. 2 and 3,  the disk is stable for $\beta^{-1}=0$,  but may
become unstable for any finite plasma beta $\beta^{-1}>0$.
 Since
${\hat\xi}^{\pm}$ are inversely proportional  to the derivative of
 the inertial moment density $\partial_r N_m$
(see the second relation (28b)  resolved with respect to
${\hat\xi}$), the vertical asymptotes I in Figs. 4, 5 arise if
$\partial_r N_m$=0 at the reference point $r=1$. As seen in Figs.
4 and 5, both positive- and negative-$g$ modes are subdivided into
two sub-modes 1 and 2 separated by a vertical asymptote I, where
$\partial_r N_m$=0, while the vertical asymptote II, where $q=0$,
bounds the region of instability ($q<0$ according to relations
(29)) of the sub-mode 2.

Figure 6 and estimations (24) demonstrate for typical values of
the disk parameters that the Hall instability is excited with
inverse growth rates of the order of the rotation period. Note
that although the angular velocity of the Keplerian rotation drops
out from the dimensionless stability problem for the toroidal
magnetic configuration, the Keplerian rotation velocity determines
the value of the Hall parameter (see Eqs. (6)).

  Let us estimate the minimal value
of the dimensional toroidal magnetic field that corresponds to the
onset of the Hall instability. According to the definition of
${\hat \xi}$ in Eqs. (6), (20), the minimal magnetic field that
gives rise to the Hall instability, is given by (Fig. 7):
\begin{equation}
B_*= {4\pi e n_* (G M)^{1/2}\over c} \,\, \epsilon  \alpha {\hat
\xi} r_*^{1/2},
\end{equation}
where ${\hat\xi}$ may be estimated as the characteristic threshold
value ${\hat\xi}^{-}$ that corresponds to the onset of the Hall
instability in Figs. (4) and (5).  Since $B_*$ in
 Eq. (36) is proportional to the  ionization degree
$\alpha$ and the aspect ratio of the disk $\epsilon$, the onset of
the Hall instability may occur in ionized weakly and magnetized
thin disks.

\section{Summary and conclusions}\label{sec:summ}

In the present study, a linear analysis of the stratification
driven Hall instability for thin equilibrium Keplerian disks with
an embedded purely toroidal magnetic field and hydrostatically
equilibrated density has been performed. The stability analysis
has been carried out in the short-wave local approximation for the
radial and vertical coordinates frozen at the reference point in
the midplane,  under the assumption of a small slope of the
vertical edge of the disk. The leading order asymptotic expansions
in small aspect ratio of the disk is employed in order to
construct the Hall MHD
 equilibrium of the thin Keplerian disks.  The Hall electric
potential is introduced which determines the equilibrium toroidal
magnetic field up to a free function of the inertial moment
density. The latter has been modelled as a sum of the current free
magnetic field (created by a current localized along the disk
axis) and the simplest quadratic approximation for the magnetic
field created by the electric current distributed over the
Keplerial portion of the disk.  The corresponding
 equilibrium profiles  contain a single free parameter of the
 model
  that may be expressed through the total current
  - the natural parameter of the disk.

It is shown that: (i) the radially-stratified equilibrium disk is
stable to radial, and unstable under vertical perturbations. The
instability is demonstrated by  the marginal stability surface in
the space of the local Hall and inverse plasma beta parameters, as
well as the free parameter of the
 model (Fig. 1); (ii) the sign of the free
parameter determines  qualitatively different behavior of the both
equilibrium and perturbations  (Figs. 2 - 5). In particular, for
negative values of the free parameter, the equilibrium disk
exhibits a maximum in the number density that corresponds to rings
of denser material, as well as a maximum in the magnetic field
(Fig. 3); (iii) the disk is stable for the infinite value  but may
become unstable for any finite value of plasma beta (Figs. 2 and
3); (iv) the current free configuration is shown to be stable, and
the existence of the distributed electric current is found to be
necessary in order to give rise to the Hall instability; (v) the
inverse growth rate is of the order of the rotation period taken
as the characteristic time scale (Fig. 6); (vi) the density
 inhomogeneity length is of the order of the characteristic disk
thickness;  (vii) the onset of the Hall instability is possible in
 thin weakly magnetized and ionized disks (Fig. 7).

 \acknowledgments

This work has been supported by the Israel Science Foundation
under Contract No. 265/00.

\clearpage

\begin{figure}
\epsscale{1.0} \plotone{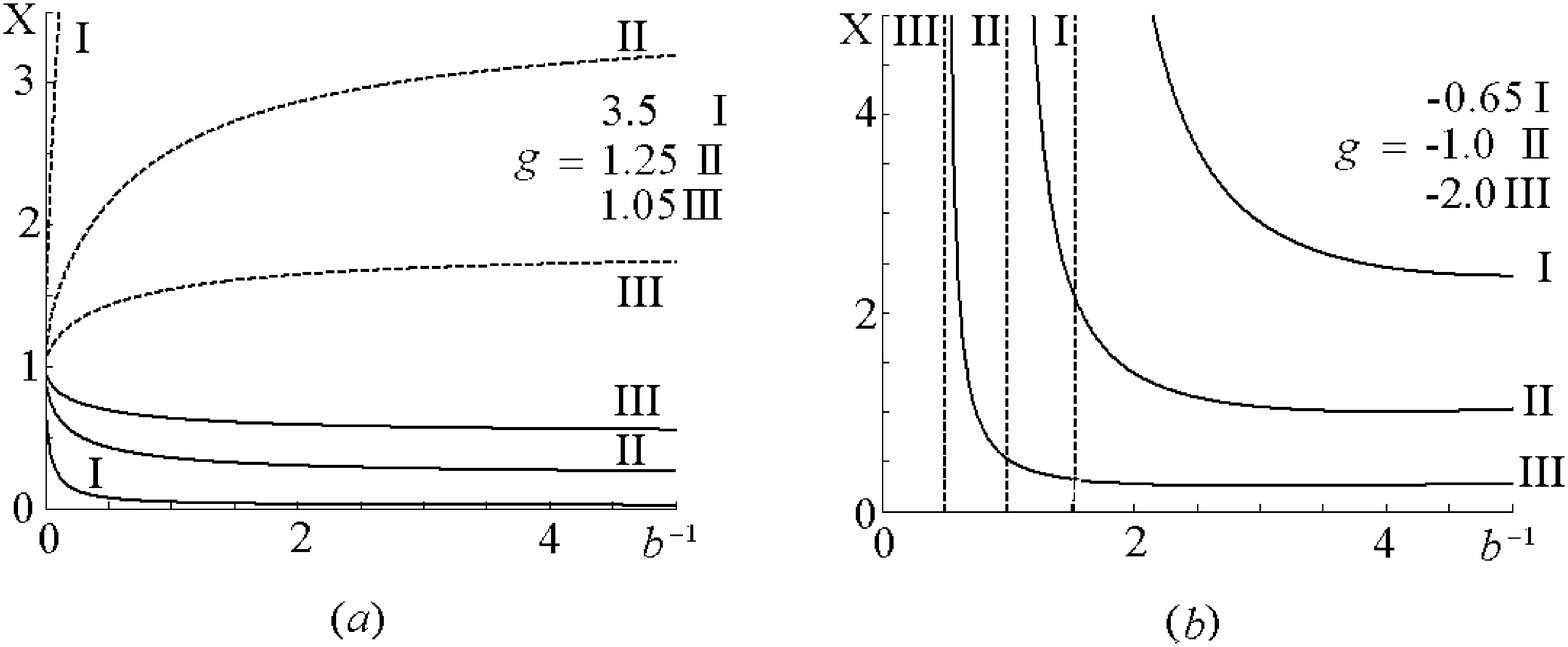} \caption{ The upper/bottom
 branches ( dotted/solid lines $X^+/X^-$) of the marginal
stability curves in the plane of the local Hall and inverse plasma
beta parameters $b^{-1}-X $ for the reference value of radial
coordinate $r=1$ in the midplane $z=0$ (curves I, II, III for
three different values $g\equiv g_0$). Instability region lies
either (a) between the upper
 and bottom  branches
  ($g>1$ and $q>1$)  or  (b) above the bottom branch
 ($g<0$ and $q<0$),
  dashed straight lines are the vertical asymptotes $q=0$.
\label{marginalocal}}
\end{figure}

\clearpage

\begin{figure}
\epsscale{1.0} \plotone{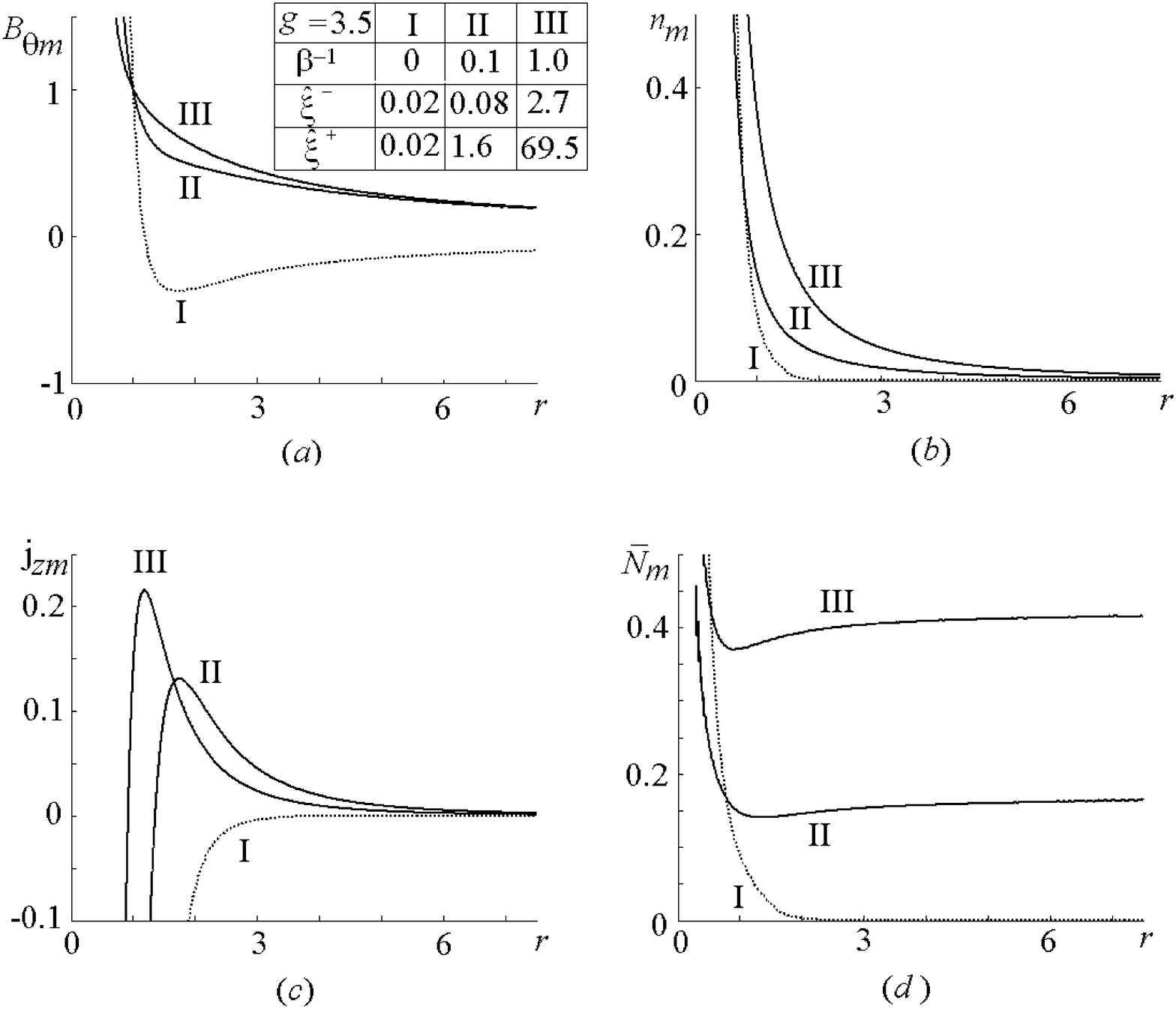} \caption{Equilibrium
midplane profiles of the toroidal magnetic field $(a)$, number
density $(b)$, distributed electric current $(c)$
 and inertial moment
density $(d)$  vs $r$ for $g\equiv g_0=3.5$  and three values of
the plasma beta $\beta$ (curves I, II, III); ${\hat \xi}^{-} <
{\hat \xi}\equiv {\hat \xi}_0 <{\hat \xi}^{+}$ belongs to the
region of instability. \label{Profiles Positive_g_}}
\end{figure}

\clearpage

\begin{figure}
\epsscale{1.0} \plotone{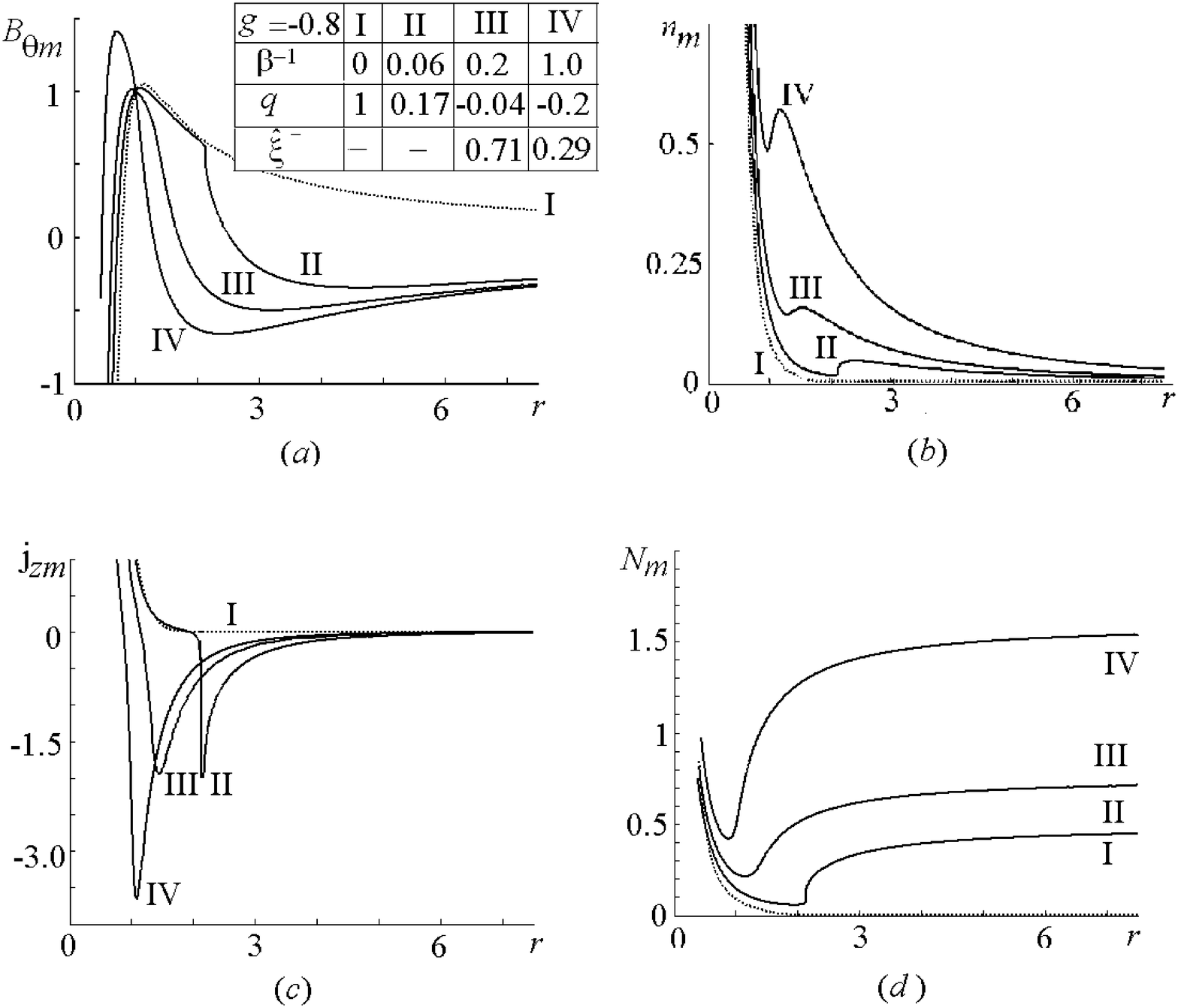} \caption{Equilibrium
midplane profiles of the toroidal magnetic field $(a)$, number
density $(b)$, distributed electric current $(c)$
 and inertial moment
density $(d)$  vs $r$ for  $g\equiv g_0=-0.8$  and three values of
the plasma beta $\beta$ (curves I, II, III);  ${\hat \xi}^{-}<
{\hat \xi}\equiv {\hat \xi}_0 $ belongs to the region of
instability. \label{Profiles Negative_g}}
\end{figure}

\clearpage

\begin{figure}
\epsscale{1.0} \plotone{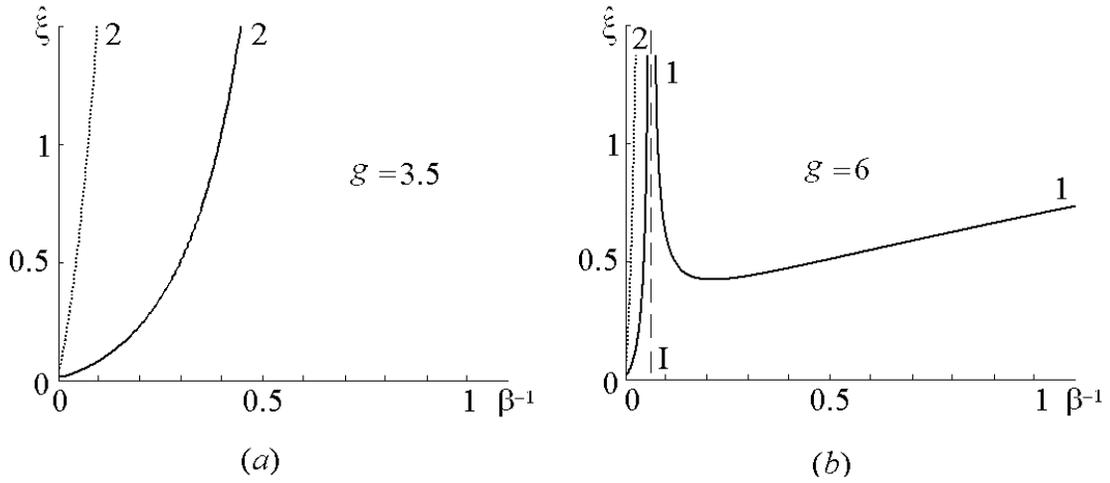} \caption{  Marginal stability
curves in the plane of the Hall criterion and inverse plasma beta
$ \beta^{-1}-{\hat \xi}$ for (a) $g=3.5$ and (b) $g=6$ ($g\equiv
g_0>1$, ${\hat \xi}\equiv {\hat \xi}_0 $,
 reference radius $r=1$ in the midplane $z=0$).
Dotted/solid lines depict the upper/bottom branches ${\hat \xi}=
{\hat \xi}^{\pm}$ of the marginal stability curves. Dashed lines
depict the asymptote I ($\partial_r N_m=0$) that separates the
sub-modes 1 and 2. The upper branch in Fig. 4b and both branches
in Fig. 4a for the  sub-mode 1 are not presented since they
correspond to the Hall parameter ${\hat \xi}>1.5$.
 \label{Marginal SC
Positive_g_0}}
\end{figure}

\clearpage

\begin{figure}
\epsscale{1.0} \plotone{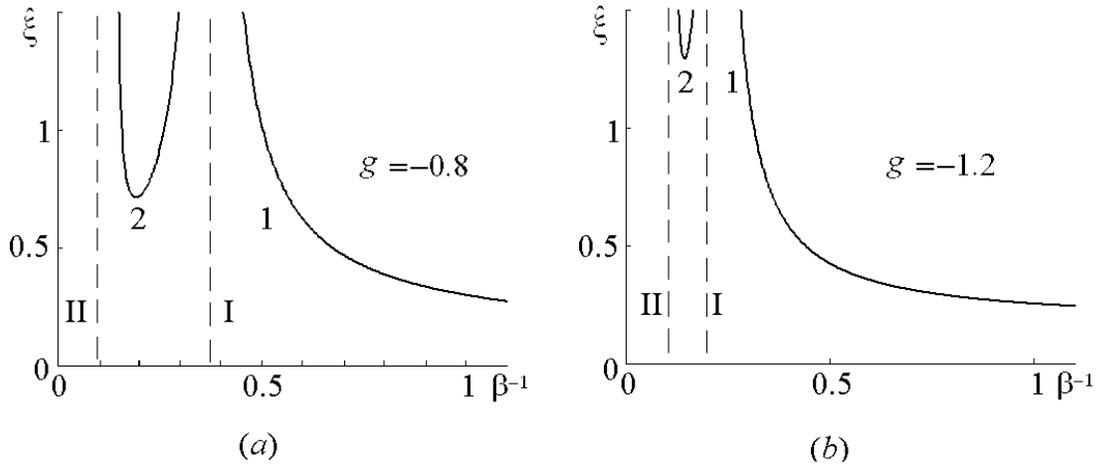} \caption{ Marginal stability
curves in the plane of the Hall criterion and inverse plasma beta
 $ \beta^{-1}-{\hat \xi}$ for (a) $g=-0.8$ and (b) $g=-1.2$
 ( $g\equiv g_0<0$, ${\hat \xi}\equiv {\hat \xi_0}$,
 reference  radius $r=1$ in the midplane $z=0$). Solid
lines depict the marginal stability curves. Dashed lines depict
the asymptote I ($\partial_r N_m=0$) that separate the sub-modes 1
and 2, and the asymptote II ($q=0$) which restricts the region of
instability for the sub-modes  2. \label{Marginal SC Negative_g}}
\end{figure}

\clearpage

\begin{figure}
\epsscale{1.0} \plotone{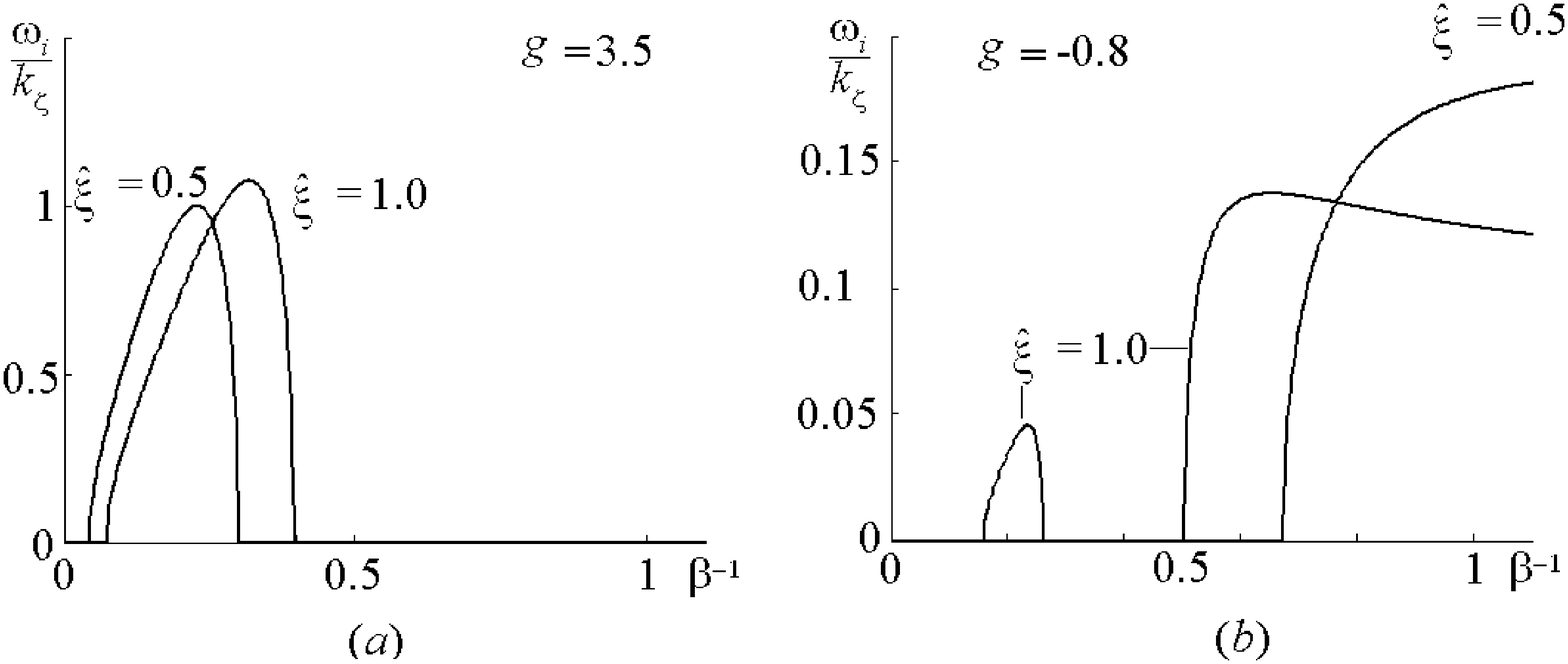} \caption{ Scaled growth rate
vs inverse plasma beta for (a) $g=3.5$ ($q>1$) and (b) $g=-0.8$
($q<0$); $g\equiv g_0$, $\,{\hat \xi}\equiv {\hat \xi}_0 $.
\label{Amplification}}
\end{figure}

\clearpage

\begin{figure}
\epsscale{0.50} \plotone{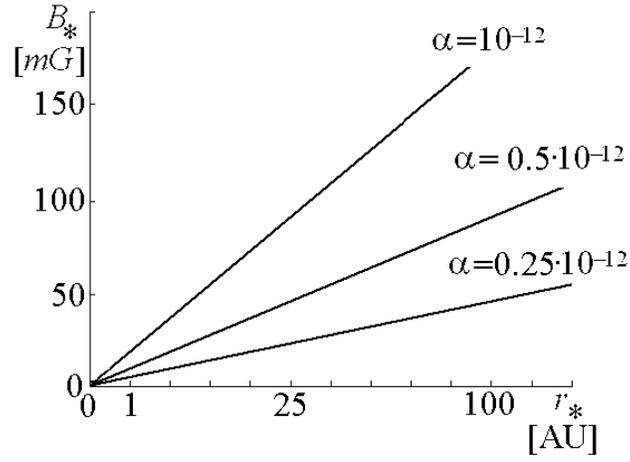} \caption{ Dimensional
characteristic toroidal magnetic field in milli-Gauss vs reference
radius in astronomical units for three values of the ionization
degree $\alpha$; ${\hat \xi}\equiv {\hat \xi_0}=0.2$;
$\epsilon=0.02$ ; $n_*\equiv n_n=10^{12}cm^{-3}$; $GM=1.3\cdot
10^{-12}cm^3/s^2$  . \label{Amplification}}
\end{figure}

\end{document}